\documentclass[aps,prl,floatfix,twocolumn]{revtex4}
\usepackage{amsmath}
\usepackage{graphicx}
\usepackage{epstopdf}

\begin{document}

\title{Experimentally generating and tuning robust entanglement between photonic qubits}

\author{B. P. Lanyon$^{1,*}$ and N. K. Langford$^{1,2}$}
\affiliation{$^1$ Department of Physics and Centre for Quantum Computer Technology, 
University of Queensland, QLD 4072, Brisbane, Australia.\\
$^2$  Faculty of Physics, University of Vienna, 
Boltzmanngasse 5, 1090, Vienna, Austria}

\begin{abstract}

We generate and study the entanglement properties of novel states composed of three polarisation-encoded photonic qubits.  By varying a single experimental parameter we can coherently move from a fully separable state to a maximally robust W state, while at all times preserving an optimally robust, symmetric entanglement configuration. We achieve a high fidelity with these configurations experimentally, including the highest reported W state fidelity. 

\end{abstract}

\maketitle

Large multipartite entangled states play a central role in many active areas of  research including quantum information, computation, communication and metrology \cite{citeulike:541803, VittorioGiovannetti11192004, PhysRevLett.86.5188}. However, while entanglement in bipartite quantum systems is well understood, multipartite entanglement is relatively unexplored and offers a far more complex structure; there are various types of entanglement that present significant generation, manipulation and characterisation challenges.  There has already been much theoretical work devoted to classifying and quantifying to what degree and in which way multipartite states are entangled \cite{PhysRevA.62.062314, verstraete-2002-65, monitoring}. Recently, experimentalists are beginning to achieve the level of control over quantum systems required to generate and study multipartite entanglement \cite{kiesel:063604, walther-2005-94, Lu:2007lr}.

In this paper we explore robust entanglement between three qubits; the simplest systems in which the phenomenon can be observed. This feature is best exemplified by the well-known GHZ and W states, which are the canonical examples of inequivalent classes of multi-qubit entanglement \cite{footnote-SLOCC}. The entanglement in a GHZ state is maximally fragile; loss of information about any single qubit leaves the remaining two in a fully separable state.  Conversely, the entanglement in a three-qubit W state is robust \cite{PhysRevA.62.062314};  loss of the information in any single qubit leaves the remaining two in an entangled state.  The question of entanglement robustness arises naturally in experimental situations from decoherence mechanisms involving loss of qubits or qubit information.  This is particularly important in quantum communication and computation where entanglement is a vital resource.

We generate and study the entanglement properties of novel states composed of three polarisation-encoded photonic qubits. We introduce and experimentally demonstrate a scheme to control the level of robust entanglement in this system, allowing tuning between a fully separable state and a maximally robust W state. We  show that the entangled states generated are, in an important sense, optimally robust against information loss and achieve high fidelity with the expected states in all cases.

\begin{figure}
\includegraphics[width=1\columnwidth]{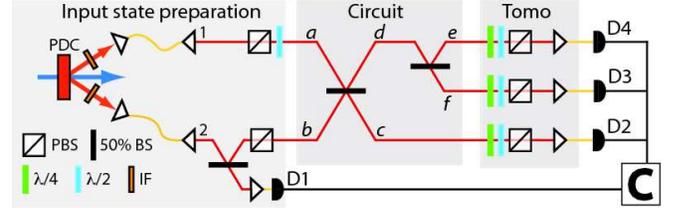}
\caption{Conceptual experimental layout. Photons are generated via SPDC of a frequency-doubled mode-locked Ti:Sapphire laser (820nm $\rightarrow$ 410nm, $\Delta \tau=80$fs at 82MHz) through a Type-I 2mm BiB$_{3}$O$_{6}$ PDC crystal. Photons are filtered by blocked interference filters (IF) at 820$\pm{1.5}$nm; collected into two single-mode optical fibers; injected into free-space modes 1 and 2; detected using fibre-coupled single photon counting modules (D1--D4). The coincident detection (C) of photons at D1--D4 selects, with high probability,  the cases of double photon-pair emission from the PDC source. With 300mW at 410nm we observe a fourfold coincidence rate of 0.1Hz.}
\vspace{-5mm}
\label{fig1:conceptual-layout}
\end{figure}

We generate our photons through double-pair emission from spontaneous parametric down conversion (SPDC), Fig.~1. Measurement of a four-fold coincidence between detectors D1--D4 selects, with high probability, the cases where the source emitted two pairs of photons into optical modes 1~\&~2.  The polarisation of two photons in the same spatio-temporal mode represents a three-level quantum system, a biphotonic qutrit \cite{BogdanovYI2003a}, with logical basis states:  $|\mathbf{0}_3\rangle{\equiv} |2_H,0_V\rangle$, $|\mathbf{1}_3\rangle{\equiv}|1_H,1_V\rangle$ and $|\mathbf{2}_3\rangle{\equiv} |0_H,2_V\rangle$.
Passing the two-photon state of mode 1  through a horizontal polarizer prepares the state $|\mathbf{0}_3\rangle$, we then create a superposition in mode $a$, using a half-wave plate set at an angle $\theta$, of the form:
\begin{equation}
\cos^22\theta |\mathbf{0}_{3}\rangle{+}\sqrt{2} \cos2\theta \sin2\theta |\mathbf{1}_{3}\rangle{+}\sin^22\theta |\mathbf{2}_{3}\rangle
\end{equation}
Mode 2 is passed to a 50\% beam splitter; detection of a single photon at D1 heralds the presence of a single photon in mode $b$; which is passed through a polarising beam splitter to prepare a polarisation qubit ($|\mathbf{0}_{2}\rangle {\equiv} |1_{H},0_{V} \rangle, |\mathbf{1}_{2}\rangle {\equiv} |0_{H},1_{V} \rangle$) in the logical state $|\mathbf{0}_{2}\rangle$. Thus a qubit and qutrit arrive simultaneously at the first 50\% beam splitter in our optical circuit.  

A successful coincidence measurement heralds the cases where a  biphotonic qutrit exits the central splitter in mode $d$ and  splits into single photon states in modes $e$ and $f$ after the final 50\% beam splitter. At the output of the circuit we find the following three-qubit joint state across modes $c$, $e$ and $f$:
\begin{eqnarray}
&&\frac{\cos^2 2\theta}{2\sqrt{2}} \;|\mathbf{0}_2,\mathbf{0}_2,\mathbf{0}_2\rangle{+}\cos 2\theta \sin 2\theta   \;|\mathbf{1}_2,\mathbf{0}_2,\mathbf{0}_2\rangle
\nonumber \\
&&{+}\;\frac{\sin^2 2\theta}{2\sqrt{2}}\;(|\mathbf{1}_2,\mathbf{1}_2,\mathbf{0}_2\rangle{+}|\mathbf{1}_2,\mathbf{0}_2,\mathbf{1}_2\rangle{-}|\mathbf{0}_2,\mathbf{1}_2,\mathbf{1}_2\rangle).
\end{eqnarray}
This is a superposition of a separable state (first two terms) and an entangled W state (last three terms).
Choosing $\theta{=}\pi/4$ injects a biphoton composed of two vertically polarised photons ($|\mathbf{2}_3\rangle$) into mode $a$ (Eq.~1) and results in a three-qubit W state at the output with probability 1/8 (Eq.~2).  Choosing  $\theta{=}0$ injects a biphoton composed of two horizontal photons ($|\mathbf{0}_3\rangle$) into mode $a$ and produces a separable state at the output, of the form $|\mathbf{0}_2,\mathbf{0}_2,\mathbf{0}_2\rangle$. 

Quantifying the amount of genuine tri-partite entanglement in a three-qubit pure state is non-trivial. The three-tangle ($\tau_3$) \cite{PhysRevA.61.052306, 3-tangle} quantifies GHZ-class entanglement and, since it is always zero for the W class \cite{PhysRevA.62.062314}, can be used to distinguish the W and GHZ classes. Using this measure combined with the technique of  \cite{PhysRevA.62.062314}, it is straightforward to show that our state (Eq.~2) belongs to the W class. An entanglement monotone useful for quantifying W-class entanglement is the tripartite negativity ($N_{3}$) \cite{sabin-2007, love-2007, Nabc} . Under this scheme  the three-qubit W state has a near maximal value of $N_{3}{=}0.94$. Quantifying how robust the entanglement in our three-qubit system is to loss requires a measure of the residual bipartite entanglement left in the two-qubit subsystem after loss of the information contained in qubit $k$ ($\rho_{\it ij}{=}\text{Tr}_{\it k}(\rho_{\it ijk})$). We choose to use the tangle ($\tau_2$) \cite{PhysRevA.61.052306}. 

By rotating $\theta$, we can tune both the the tripartite ($N_3$) and residual bipartite ($\tau_2$) entanglement between zero and that of a W state. Interestingly, it is straightforward to show that as we tune, the residual bipartite entanglement always remains symmetrically distributed between each pair of qubits, i.e. $\tau_2(\rho_{ce}){=}\tau_2(\rho_{cf}){=}\tau_2(\rho_{fe}){=}4 sin^4\theta/(cos2\theta - 2)^2$. As a result the amount of entanglement left in the subsystem is, for all $\theta$, \emph{independent} of which qubit is lost. 

\begin{figure}
\includegraphics[width=1\columnwidth]{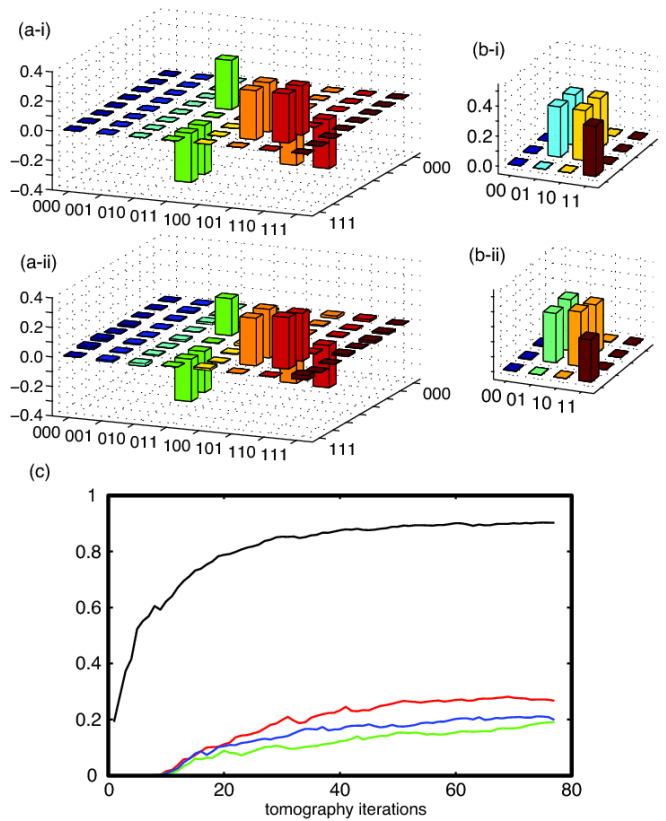}
\vspace{-6mm}
\caption{Results for $\theta{=}\pi/4$ (Eq.~2). (a-i) Ideal and (a-ii) measured three-qubit density matrices. We find a fidelity of $0.90{\pm}{0.03}$ 
with the ideal W state, a linear entropy of $S_L{=}0.20{\pm}{0.03}$ 
and a tripartite negativity of $N_3{=}0.80{\pm}{0.03}$ \cite{Fid-SL}. 
(b-i) Ideal and (b-ii) measured reduced state of qubits $c$ and $e$ reconstructed via $\rho_{\it ef}{=}\text{Tr}_{\it c}(\rho_{\it cef})$. We find a fidelity with the ideal MEMS \cite{PhysRevA.64.030302, peters:133601} of $0.94{\pm}{0.02}$, a linear entropy of $S_L{=}0.61{\pm}0.02$ (ideal 5/9) and a tangle of $\tau_2{=}0.27{\pm}0.03$ (ideal 4/9). c) Iterative tomography results for the measured states shown in a)-b). A black line shows the development of the three-qubit state fidelity. Red, blue and green lines show the development of the tangle ($\tau_2$) in the reduced states $\rho_{\it ef}$, $\rho_{\it ec}$ and $\rho_{\it fc}$, respectively.
\vspace{-5mm}
}
\label{fig2:wstate}
\end{figure}

We measure three-qubit output states using over-complete polarisation tomography of modes $c$, $e$ and $f$, performing 216 separate measurements \cite{lanyon-2007} in four-fold coincidence between detectors D1--D4.  With four-fold detection rates of approximately 0.1$s^{-1}$ we measure for several days to acquire sufficient counts for an accurate reconstruction.  Instead of performing a single measurement set over this time we take many shorter sets, each taking around 80 minutes.  This \emph{iterative} tomography technique provides many advantages.  Most importantly, a complete reconstruction of the density matrix  is possible after each iteration, allowing us to analyse how our estimates of state properties are developing througout the measurement process.  This allows diagnosis of serious practical problems, such as time-dependant optical misalignment, far earlier than would otherwise be possible.  Shorter measurement sets are also less prone to errors introduced by certain fluctuations in the optical source brightness.  We do not use an iteration time of less than 80 minutes, because the fixed time required to rotate the polarisation analysis plates begins to significantly impinge on the photon counting time. We use convex optimisation and fixed weight estimation to reconstruct physical density matrices and Monte-Carlo simulations of Poissonian photon counting fluctuations for error analysis \cite{obrien:080502, wellwhereisit, LangfordNK2007phd}.

In practice, our beamsplitters are imperfect, imparting systematic unitary operations on the optical modes. While the entanglement properties of our states are not affected by these local operations, state fidelities are. For simplicity, we corrected for these effects numerically, alternatively such unitaries could be corrected using standard waveplates.

Measured and ideal three-qubit density matrices for $\theta{=}\pi/4$ (Eq.~2) are shown in Fig~2a). We find the highest-reported fidelity with the ideal W state and a low linear entropy (see caption). The state fidelity violates the entanglement witness for a W state by 7 standard deviations \cite{bourennane:087902} and has a tri-partite negativity of $N_{3}\,{=}\,0.80{\pm}0.03$. Fig.~2b shows the reduced states of qubits $e$ and $f$, calculated by numerical application of a partial trace to the states in Fig~2a) ($\rho_{\it ef}\,{=}\,\text{Tr}_{\it c}\{\rho_{\it cef}\}$). We find a high fidelity with the ideal maximally entangled mixed state (MEMS) \cite{PhysRevA.64.030302, peters:133601} (see caption). This state is unambiguously entangled with a  tangle of $\tau_2{=}0.27{\pm}0.03$ (ideal: 4/9), demonstrating the robustness of the entanglement in the three-qubit state to loss.  

Fig.~2c shows how our knowledge of key properties of the reconstructed states (Fig.~2a-b) developed over the iterative measurement process. The asymptotic trends show that we measured for a sufficient period of time such that our reconstructed states are a fair representation of the generated states. Note that the robust entanglement in the reduced two-qubit states does not become apparent until after several iterations.

Fig.~3 shows experimental results for a range of $\theta$ (Eq.~2). Fig.~3a compares the measured and ideal tripartite and bipartite entanglement distributions in the three-qubit output states ($\rho_{\it cef}$).  We find high fidelities with the ideal symmetric robust three-qubit states, as detailed in the figure caption. 

We also measure the reduced two-qubit states \emph{directly} by removing the polarisation analysis optics from one qubit output mode at a time and only detecting its presence as a trigger---physically realising the loss of qubit information. This was repeated for each qubit to test the symmetry of our measured states. Besides offering an unambiguous demonstration of robust entanglement, this approach offers an increased count-rate over that observed when measuring three-qubit states. 
We perform over-complete polarisation tomography of the remaining two qubits using 36 separate measurements \cite{langford:210504}. 
Fig.~3b presents the results plotted on the tangle vs linear entropy plane \cite{PhysRevA.64.030302}. 
The dashed line shows the path of the ideal reduced states for varying $\theta$ (Eq.~2); the residual tangle increases linearly with the entropy, with the pure separable state for $\theta{=}0$ at the origin, and a MEMS for $\theta{=}\pi/4$ (corresponding to the ideal reduced state of the three-qubit W state shown in Fig.~1b-i)).  
Due to the symmetry properties of the ideal three-qubit states, this trend does not depend on which qubit is lost. The results show a good correlation with the ideal trend and high fidelities with the expected states (see figure caption);  we can tune the level of robust entanglement in our system. 

The main discrepancies in experimental results are an increased mixture and correspondingly decreased entanglement in the measured states---common errors in optical quantum circuits.  These effects result from a combination of higher order emissions from SPDC \cite{Tills} and the photonic qubits entangling to undesired, unmeasured degrees of freedom, such as time and spatial mode \cite{rohde-2006-73}.

\begin{figure}
\includegraphics[width=1\columnwidth]{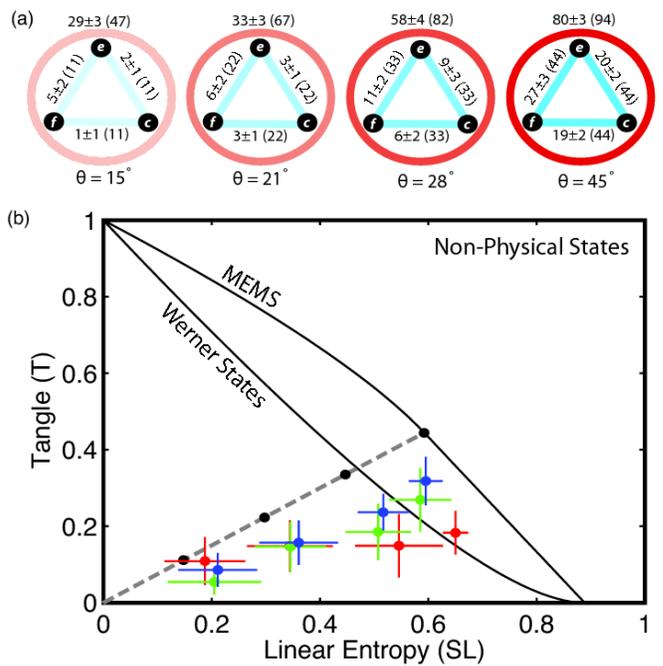}\
\caption{Key properties of measured states, for $\theta{=}\{15^{\circ},  21^{\circ}, 28^{\circ},  45^{\circ}\}$, Eq.~2. (a) Measured (ideal) entanglement distributions (in \%) in three-qubit output states ($\rho_{\it cef}$). Labelled black circles represent qubits, a red circle represents tripartite entanglement ($N_{3}$) and a blue line between qubits $i$ and $j$ represents bipartite (robust) entanglement ($\tau_2$) in the reduced state ($\rho_{\it ij}{=}\text{Tr}_{\it k}\{\rho_{\it ijk}\}$) \cite{sabin-2007}. We find high fidelities with the ideal configurations of 
$\{0.90{\pm}{0.02}, 0.84{\pm}{0.03}, 0.84{\pm}{0.05}, 0.90{\pm}{0.03}\}$  
and low linear entropies 
$\{0.20{\pm}{0.03},0.22{\pm}{0.03}, 0.25{\pm}{0.03}, 0.20{\pm}{0.03}\}$, 
respectively.  (b) Tangle vs linear entropy plane \cite{PhysRevA.64.030302} showing results for reduced two-qubit states, $\rho_{\it ij}$, measured \emph{directly} by removing the polarisation analysis optics of qubit $k$, and performing two-qubit tomography. Results for loss of qubit $c$, $e$ and $f$ are shown in blue, red and green, respectively (ideal cases shown in black).
The dashed line shows the ideal trend calculated from Eq.~2. The Werner states \cite{PhysRevA.40.4277} and MEMS \cite{PhysRevA.64.030302, peters:133601} are also shown. The average fidelity of the measured reduced two-qubit states with the ideal is $0.97{\pm}{0.02}$.
\vspace{-5mm}
}
\label{fig3:tunable-robustness}
\end{figure}

D\"ur and Vidal showed that entanglement in a three-qubit W state is maximally robust in two respects \cite{PhysRevA.62.062314}. Firstly, it maximises the ``weakest link'' residual tangle between two-qubit subsystems, namely: 
\begin{equation}
\tau_{\rm 2,min}(\rho_{\it abc}) = \min \; \{ \tau_2(\rho_{\it ab}), \tau_2(\rho_{\it ac}), \tau_2(\rho_{\it bc}) \},
\end{equation}
where $\rho_{ij}{=}\text{Tr}_{\it{k}}\{\rho_{\it ijk}\}$ is the reduced state of $\rho_{\it abc}$ after the loss of qubit~$k$.
Secondly, it has the highest average residual tangle, namely optimising the function:
\begin{equation}
\overline{\tau_2}(\rho_{\it abc}) = \frac{1}{3} \; (\tau_2(\rho_{\it ab})+\tau_2(\rho_{\it ac})+\tau_2(\rho_{\it bc})).
\end{equation}
Fig.~4 shows $N_{3}$ versus $\tau_{\rm 2,min}$ for 300,000 pure three-qubit states randomly selected using the Haar measure \cite{Haar}, with the colourmap representing the corresponding three-tangle ($\tau_3$). The black line shows the curve for our ideal states (Eq.~2), from the separable state at the origin ($\theta{=}0$) to the W state ($\theta{=}\pi/4$), which reaches the maximum possible $\tau_{\rm 2,min}$ value of 4/9.  This clearly represents a boundary in robust configurations of entanglement:  for a given level of genuine pure-state three-qubit entanglement ($N_{3}$) the weakest bipartite link between any pair of qubits in our states is of optimal strength.  States that are not optimal in this sense have at least one weaker bipartite link: there is a `linchpin' qubit which, if lost, will leave less bipartite entanglement between the remaining qubits.  Note that the density of states near the boundary described by our states is lower because the set of three-qubit W-class states is of measure zero compared with the set of three-qubit GHZ-class states \cite{PhysRevA.62.062314}.  Fig.~4 includes the positions of the four measured states shown in Fig.~3a. Note that, even though our measured W state has a fidelity of over $90\%$ with the ideal, the value of $\tau_{\rm 2,min}$ is less than half of the expected value. Clearly maximising this property is far more experimentally challenging than achieving a high state fidelity. 

Similar numerical simulations show that our states are not optimal under the criteria of Eq.~4; there are states which retain a higher average residual entanglement. However, states that improve on ours in this respect do so at the expense of losing a symmetric distribution of entanglement; they always have at least one weaker bipartite link. An extreme example is the state $|\mathbf{0},\psi^+\rangle$, where $\psi^+$ is a maximally entangled Bell state, which has $N_3{=}0$ and an average subsystem tangle of $\overline{\tau_2}{=}0.33$.  However loss of either qubit in the Bell state is sufficient to leave a fully separable state, so $\tau_{\rm 2,min}{=}0$.  (For $N_3{=}0$, our states have $\tau_{\rm 2,min}{=}0$ and $\overline{\tau_2}{=}0$.)

\begin{figure}[h]
\includegraphics[width=1\columnwidth]{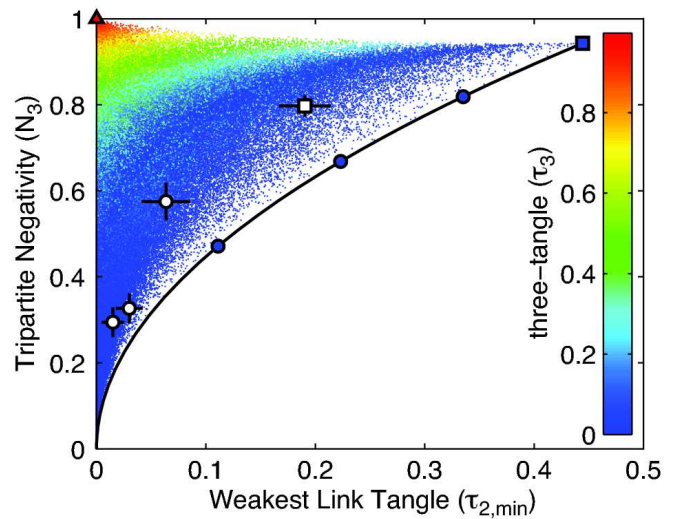}
\vspace{-8mm}
\caption{Plot of $N_{3}$ versus $\tau_{\rm 2,min}$ (Eq.~3) \cite{Nabc}. The positions of 300,000 randomly selected pure  three-qubit states are shown \cite{Haar}, with the colour map representing the corresponding three-tangle ($\tau_{3}$). The black curve shows the ideal positions of our states (Eq.~2) as we vary $\theta$ from $0$ (at the origin) to $\pi/4$. Results for the four experimentally measured three-qubit states are shown (Fig.~3a). The three-qubit GHZ (red triangle) and W state (blue square) are included. }
\vspace{-7mm}
\label{fig4:optimal-robustness}
\end{figure}

In conclusion, we have experimentally demonstrated control over the level of robust entanglement between three photonic qubits.  In the ideal case our technique allows tuning from a fully separable state to a maximally robust W state, whilst maintaining a symmetric distribution of bipartite entanglement between each pair of qubits. Furthermore, in doing so we maintain, in an important sense, an optimal configuration of robust entanglement.  We achieve high state fidelities with these configurations and demonstrate control over the three-qubit W class of entanglement. 

We wish to thank  Guifre Vidal and Andrew White for valuable discussions. This work was supported by the Australian Research Council and DEST Endeavor programs.

{\small $^*$Electronic address: {\texttt lanyon@physics.uq.edu.au}}


\begin{thebibliography}{27}
\expandafter\ifx\csname natexlab\endcsname\relax\def\natexlab#1{#1}\fi
\expandafter\ifx\csname bibnamefont\endcsname\relax
  \def\bibnamefont#1{#1}\fi
\expandafter\ifx\csname bibfnamefont\endcsname\relax
  \def\bibfnamefont#1{#1}\fi
\expandafter\ifx\csname citenamefont\endcsname\relax
  \def\citenamefont#1{#1}\fi
\expandafter\ifx\csname url\endcsname\relax
  \def\url#1{\texttt{#1}}\fi
\expandafter\ifx\csname urlprefix\endcsname\relax\def\urlprefix{URL }\fi
\providecommand{\bibinfo}[2]{#2}
\providecommand{\eprint}[2][]{\url{#2}}

\bibitem[{\citenamefont{Nielsen and Chuang}(2000)}]{citeulike:541803}
\bibinfo{author}{M.~A. Nielsen and I. L. Chuang},
  \emph{\bibinfo{title}{Quantum Computation and Quantum Information}}
  (\bibinfo{publisher}{{Camb. Univ. Press}}, \bibinfo{year}{2000})

\bibitem[{\citenamefont{Giovannetti et~al.}(2004)\citenamefont{Giovannetti,
  Lloyd, and Maccone}}]{VittorioGiovannetti11192004}
\bibinfo{author}{\bibfnamefont{V.}~\bibnamefont{Giovannetti}},
  \bibinfo{author}{\bibfnamefont{S.}~\bibnamefont{Lloyd}}, \bibnamefont{and}
  \bibinfo{author}{\bibfnamefont{L.}~\bibnamefont{Maccone}},
  \bibinfo{journal}{Science} \textbf{\bibinfo{volume}{306}},
  \bibinfo{pages}{1330} (\bibinfo{year}{2004}).
  
  \bibitem{PhysRevLett.86.5188}
\bibinfo{author}{Raussendorf, R.} \& \bibinfo{author}{Briegel, H.~J.}
\newblock \bibinfo{title}{A one-way quantum computer}.
\newblock \emph{\bibinfo{journal}{Phys. Rev. Lett.}}
  \textbf{\bibinfo{volume}{86}}, \bibinfo{pages}{5188--5191}
  (\bibinfo{year}{2001}).

\bibitem[{\citenamefont{D\"ur et~al.}(2000)\citenamefont{D\"ur, Vidal, and
  Cirac}}]{PhysRevA.62.062314}
\bibinfo{author}{\bibfnamefont{W.}~\bibnamefont{D\"ur}},
  \bibinfo{author}{\bibfnamefont{G.}~\bibnamefont{Vidal}}, \bibnamefont{and}
  \bibinfo{author}{\bibfnamefont{J.~I.} \bibnamefont{Cirac}},
  \bibinfo{journal}{Phys. Rev. A} \textbf{\bibinfo{volume}{62}},
  \bibinfo{pages}{062314} (\bibinfo{year}{2000}).

\bibitem[{\citenamefont{Verstraete et~al.}(2002)\citenamefont{Verstraete,
  Dehaene, {De Moor}, and Verschelde}}]{verstraete-2002-65}
\bibinfo{author}{F. Verstraete \emph{et al.}},
  \bibinfo{journal}{Phys. Rev. A} \textbf{\bibinfo{volume}{65}},
  \bibinfo{pages}{052112} (\bibinfo{year}{2002}).

\bibitem[{\citenamefont{Bastin et~al.}(2007)\citenamefont{Bastin, Thiel, {von
  Zanthier}, Lamata, Solano, and Agarwal}}]{monitoring}
\bibinfo{author}{T. Bastin \emph{et al.}}
  \bibinfo{howpublished}{arXiv.org:0710.3720}
  (\bibinfo{year}{2007}).

\bibitem[{\citenamefont{Kiesel et~al.}(2007)\citenamefont{Kiesel, Schmid,
  T\'{o}th, Solano, and Weinfurter}}]{kiesel:063604}
\bibinfo{author}{N. Kiesel \emph{et al.}},
  \bibinfo{journal}{Phys. Rev. Lett.} \textbf{\bibinfo{volume}{98}},
  \bibinfo{eid}{063604} (\bibinfo{year}{2007}).

\bibitem[{\citenamefont{Walther et~al.}(2005)\citenamefont{Walther, Resch, and
  Zeilinger}}]{walther-2005-94}
\bibinfo{author}{\bibfnamefont{P.}~\bibnamefont{Walther}},
  \bibinfo{author}{\bibfnamefont{K.~J.} \bibnamefont{Resch}}, \bibnamefont{and}
  \bibinfo{author}{\bibfnamefont{A.}~\bibnamefont{Zeilinger}},
  \bibinfo{journal}{Phys. Rev. Lett.} \textbf{\bibinfo{volume}{94}},
  \bibinfo{pages}{240501} (\bibinfo{year}{2005}).

\bibitem[{\citenamefont{Lu et~al.}(2007)\citenamefont{Lu, Zhou, Guhne, Gao,
  Zhang, Yuan, Goebel, Yang, and Pan}}]{Lu:2007lr}
\bibinfo{author}{C.-Y. Lu \emph{et al.}},
  \bibinfo{journal}{Nat. Phys.} \textbf{\bibinfo{volume}{3}},
  \bibinfo{pages}{91} (\bibinfo{year}{2007}).

\bibitem{footnote-SLOCC}
\bibinfo{howpublished}{The GHZ and W states belong to inequivalent entanglement classes because they are not interconvertible via stochastic local operations and classical communication (SLOCC) \cite{PhysRevA.62.062314}.  Robust entanglement, however, is variant under SLOCC and does therefore not fit within this classification scheme \cite{walther-2005-94}.}

\bibitem[{\citenamefont{Bogdanov et~al.}(2003)\citenamefont{Bogdanov,
  Krivitsky, and Kulik}}]{BogdanovYI2003a}
\bibinfo{author}{\bibfnamefont{Y.~I.} \bibnamefont{Bogdanov}},
  \bibinfo{author}{\bibfnamefont{L.~A.} \bibnamefont{Krivitsky}},
  \bibnamefont{and} \bibinfo{author}{\bibfnamefont{S.~P.} \bibnamefont{Kulik}},
  \bibinfo{journal}{J. Exp. Theor. Phys. Lett.} \textbf{\bibinfo{volume}{78}},
  \bibinfo{pages}{352} (\bibinfo{year}{2003}).

\bibitem[{\citenamefont{Coffman et~al.}(2000)\citenamefont{Coffman, Kundu, and
  Wootters}}]{PhysRevA.61.052306}
\bibinfo{author}{\bibfnamefont{V.}~\bibnamefont{Coffman}},
  \bibinfo{author}{\bibfnamefont{J.}~\bibnamefont{Kundu}}, \bibnamefont{and}
  \bibinfo{author}{\bibfnamefont{W.~K.} \bibnamefont{Wootters}},
  \bibinfo{journal}{Phys. Rev. A} \textbf{\bibinfo{volume}{61}},
  \bibinfo{pages}{052306} (\bibinfo{year}{2000}).

\bibitem[{3-t()}]{3-tangle}
\bibinfo{howpublished}{The three-tangle of a pure three-qubit state
  $\rho_{abc}$ is defined as $\tau_{3}= 4 \det\rho_{a}{-}C_{ab}{-}C_{bc}$,
  where $C_{ij}$ is the concurrance of the reduced state $\rho_{ij}$
  \cite{PhysRevA.61.052306}.}

\bibitem[{\citenamefont{Sabin and Garcia-Alcaine}(2007)}]{sabin-2007}
\bibinfo{author}{\bibfnamefont{C.}~\bibnamefont{Sabin}} \bibnamefont{and}
  \bibinfo{author}{\bibfnamefont{G.}~\bibnamefont{Garcia-Alcaine}},
  \bibinfo{journal}{arXiv:0707.1780}  (\bibinfo{year}{2007}).

\bibitem[{\citenamefont{Love}(2007)}]{love-2007}
\bibinfo{author}{\bibfnamefont{P.~J.~e.} \bibnamefont{Love}},
  \bibinfo{journal}{Quant. Info. Proc.} \textbf{\bibinfo{volume}{6}}
  (\bibinfo{year}{2007}).

\bibitem[{Nab()}]{Nabc}
\bibinfo{howpublished}{The multiplicative tripartite negativity of a
  three-qubit state is defined as
  $N_{3}{=}(N_{a(bc)}N_{b(ac)}N_{c(ab)})^{\frac{1}{3}}$, where the bipartite
  negativities $N_{j(kl)}{=}\emph{max}\{0,-2\sum_i \lambda_{i}^{j}\}$ where
  $\lambda_{i}^{j}$ are the negative eigenvalues of the partial transpose of
  the total state with respect to the subsystem~$j$ \cite{sabin-2007}.}

\bibitem[{Fid()}]{Fid-SL}
\bibinfo{howpublished}{Fidelity is
  $F(\rho,\sigma){\equiv}\mathrm{Tr}[\sqrt{\sqrt{\rho} \sigma
  \sqrt{\rho}}]^{2}$; linear entropy is $S_{L} {\equiv}$ $d
  (1{-}\mathrm{Tr}[\rho^2])/(d{-}1)$, where $d$ is the state dimension
  \cite{langford:210504}.}

\bibitem[{\citenamefont{Munro et~al.}(2001)\citenamefont{Munro, James, White,
  and Kwiat}}]{PhysRevA.64.030302}
\bibinfo{author}{W.~J. Munro \emph{et al.}},
  \bibinfo{journal}{Phys. Rev. A} \textbf{\bibinfo{volume}{64}},
  \bibinfo{pages}{030302} (\bibinfo{year}{2001}).

\bibitem[{\citenamefont{Peters et~al.}(2004)\citenamefont{Peters, Altepeter,
  Branning, Jeffrey, Wei, and Kwiat}}]{peters:133601}
\bibinfo{author}{N.~A. Peters \emph{et al.}},
  \bibinfo{journal}{Phys. Rev. Lett.} \textbf{\bibinfo{volume}{92}},
  \bibinfo{eid}{133601} (\bibinfo{year}{2004}).

\bibitem[{\citenamefont{Lanyon et~al.}(2007)\citenamefont{Lanyon, Weinhold,
  Langford, Barbieri, James, Gilchrist, and White}}]{lanyon-2007}
\bibinfo{author}{B.~P. Lanyon \emph{et al.}},
  \bibinfo{journal}{Phys. Rev. Lett.} \textbf{\bibinfo{volume}{99}},
  \bibinfo{pages}{250505} (\bibinfo{year}{2007}).

\bibitem[{\citenamefont{O'Brien et~al.}(2004)\citenamefont{O'Brien, Pryde,
  Gilchrist, James, Langford, Ralph, and White}}]{obrien:080502}
\bibinfo{author}{J.~L. O'Brien \emph{et al.}},
  \bibinfo{journal}{Phys. Rev. Lett.} \textbf{\bibinfo{volume}{93}},
  \bibinfo{eid}{080502} (\bibinfo{year}{2004}).

\bibitem[{\citenamefont{de~Burgh A.~Doherty and
  Gilchrist}(2007)}]{wellwhereisit}
\bibinfo{author}{\bibfnamefont{M.}~\bibnamefont{de~Burgh A.~Doherty}}
  \bibnamefont{and} \bibinfo{author}{\bibfnamefont{A.}~\bibnamefont{Gilchrist, in preparation}}
  (\bibinfo{year}{2007}).

\bibitem[{\citenamefont{Langford}(2007)}]{LangfordNK2007phd}
\bibinfo{author}{\bibfnamefont{N.~K.} \bibnamefont{Langford}}, Ph.D. thesis,
  \bibinfo{school}{The University of Queensland}, \bibinfo{address}{Brisbane,
  QLD, Australia} (\bibinfo{year}{2007}).

\bibitem[{\citenamefont{Bourennane et~al.}(2004)\citenamefont{Bourennane, Eibl,
  Kurtsiefer, Gaertner, Weinfurter, G\"{u}hne, Hyllus, Bru\ss, Lewenstein, and
  Sanpera}}]{bourennane:087902}
\bibinfo{author}{M. Bourennane \emph{et al.}},
  \bibinfo{journal}{Phys. Rev. Lett.} \textbf{\bibinfo{volume}{92}},
  \bibinfo{eid}{087902} (\bibinfo{year}{2004}).

\bibitem[{\citenamefont{Langford et~al.}(2005)\citenamefont{Langford, Weinhold,
  Prevedel, Resch, Gilchrist, O'Brien, Pryde, and White}}]{langford:210504}
\bibinfo{author}{N.~K. Langford \emph{et al.}},
  \bibinfo{journal}{Phys. Rev. Lett.} \textbf{\bibinfo{volume}{95}},
  \bibinfo{eid}{210504} (\bibinfo{year}{2005}).

\bibitem[{\citenamefont{Weinhold et~al.}(2007)\citenamefont{Weinhold,
  others, White}}]{Tills}
\bibinfo{author}{T.~J. Weinhold \emph{et al.} in preparation}
  (\bibinfo{year}{2007}).

\bibitem[{\citenamefont{Rohde and Ralph}(2006)}]{rohde-2006-73}
\bibinfo{author}{\bibfnamefont{P.~P.} \bibnamefont{Rohde}} \bibnamefont{and}
  \bibinfo{author}{\bibfnamefont{T.~C.} \bibnamefont{Ralph}},
  \bibinfo{journal}{Phys. Rev. A} \textbf{\bibinfo{volume}{73}},
  \bibinfo{pages}{062312} (\bibinfo{year}{2006}).

\bibitem[{\citenamefont{Werner}(1989)}]{PhysRevA.40.4277}
\bibinfo{author}{\bibfnamefont{R.~F.} \bibnamefont{Werner}},
  \bibinfo{journal}{Phys. Rev. A} \textbf{\bibinfo{volume}{40}},
  \bibinfo{pages}{4277} (\bibinfo{year}{1989}).

\bibitem[{\citenamefont{Haar}(1993)}]{Haar}
\bibinfo{author}{\bibfnamefont{A.}~\bibnamefont{Haar}}, \bibinfo{journal}{Ann.
  Math.} \textbf{\bibinfo{volume}{34}} \bibinfo{page}{147-69} (\bibinfo{year}{1993}).

\end{thebibliography}

\end{document}